\documentclass[twocolumn,letter]{jpsj2} 
\usepackage{slashbox}
\title{Is Fermi-surface nesting the origin of superconductivity 
in iron pnictides?: \\
A fluctuation-exchange-approximation study}

\author{Ryotaro \textsc{Arita}$^{1,2,3}$ and Hiroaki \textsc{Ikeda}$^4$} 
\inst{
$^1$Department of Applied Physics, University of Tokyo, Tokyo 153-0064, Japan\\
$^2$JST, TRIP, Sanbancho, Chiyoda, Tokyo 102-0075, Japan\\
$^3$JST, CREST, Hongo, Tokyo 113-8656, Japan CREST\\
$^4$Department of Physics, Kyoto Univ, Kyoto 606-8502, Japan
}
\recdate{today}

\abst{
We study whether Fermi-surface (FS) nesting can give rise to high-temperature
superconductivity in iron pnictides. 
Starting with {\em ab initio} construction of an effective
four-orbital model, we employ the fluctuation-exchange
approximation to show that FS does not necessarily favor the stripe
antiferromagnetic order observed in experiments, 
especially for realistic electronic correlations. If
superconductivity in iron pnictides is magnetically mediated and has 
fully-gapped sign-reversing
s-wave symmetry, our results suggest that the pairing interaction 
does not arise only from FS nesting and exchange 
interactions between local moments in the Fe 3d orbitals may play a crucial role.}

\kword{iron-based superconductor, magnetism, unconventional superconductivity}

\begin{document}
\maketitle

The seminal discovery of superconductivity in LaFeAsO doped 
with fluorine\cite{Kamihara} and subsequent updatings of the record of
the transition temperature ($T_c$) in
related iron pnictides and iron chalcogenites\cite{IshidaRev} have 
attracted great interest in these compounds.
Since there is close proximity of 
antiferromagnetic(AF) order and superconductivity in the phase 
diagram, the possibility of magnetically mediated superconductivity 
has been intensively studied\cite{MazinSchmalian}.

When we consider the interplay between magnetism and superconductivity,
one of the key issues to be clarified is, whether the magnetism
should be understood in terms of localized spins or itinerant electrons.
While there have been a variety of proposals and calculations for 
the so-called $J_1$-$J_2$ model based on the 
the strong coupling scenario\cite{j1j2}, 
it has also been claimed that
the weak coupling approach should work 
better\cite{KurokiPRL,KurokiPRB,Scalapino,IkedaJPSJ,FLEX,RG,Nomura}, and
the problem which picture describes the system more 
properly has been an issue of hot debates\cite{MazinSchmalian}. 

Among many studies based on the weak coupling scenario,
recently, a detailed calculation by the random phase approximation (RPA)
for a five-orbital model 
derived from {\em ab initio} calculation based on
the local density approximation (LDA)
has been performed. There, it has been discussed how changes in the 
crystal structure (i.e., the atomic configuration) affects 
superconductivity\cite{KurokiPRB}, and it has been concluded
that the weak coupling approach succeeds
to explain the qualitative tendency of pairing instability
in iron pnictides: when the pnictogen height is low (high) as 
in the case of LaFePO (LaFeAsO), low (high) $T_c$ nodal 
(nodeless) pairing is favored\cite{LaFePO}.

On the other hand, estimates of the interaction parameters 
by various {\em ab initio} methods suggest that iron pnictides
are not weakly but {\em moderately} correlated\cite{cRPA1,cRPA2,cLDA}.
Indeed, photoemission spectrum indicates that 
the mass enhancement due to the electron correlations is as 
large as 2 $\sim$ 3\cite{ARPES}, whose effect is
neglected in RPA. Since Fermi-surface (FS) nesting 
plays a crucial role in the weak coupling approach, and 
the orbital/momentum dependence of the self-energy 
correction can seriously affect the nesting condition, 
it is of great interest to see whether the weak coupling approach
based on FS nesting survives 
even if we consider the self-energy correction.

One of the standard ways to investigate the effect of the 
self-energy correction is to employ the dynamical mean 
field theory (DMFT)\cite{DMFTrev}.
While several DMFT calculations combined with LDA
have been performed\cite{Haule,cLDA,Vollhardt,Georges}, 
the recent calculation with realistic interaction 
parameters\cite{Georges} for LaFeAsO has shown that 
the orbital-dependent DMFT correction lowers
the level of $d_{x^2{\mathchar`-}y^2}$ at the $\Gamma$ point
in the Brillouin zone (hereafter, the local coordinate system
centered at Fe is such that the $x$- and $y$-axis point
towards neighboring As). As is discussed below, in such a situation,
the nesting condition is expected to be worse, which
suggests that the nesting scenario is not necessarily robust
against the self-energy correction.
 
While the self energy in DMFT is momentum independent,
the momentum dependence of the self energy is also crucial 
for FS nesting. The purpose of the 
present study is to investigate the effect of momentum-dependent 
self energy for realistic band structure and 
electronic correlations by means of the fluctuation-exchange
approximation(FLEX)\cite{Bickers}.
Although FLEX can become problematic to address some high energy 
features (e.g., it fails to describe a Mott insulator), 
FLEX is generally expected to be reliable 
for description of low energy physics originating from
FS nesting\cite{Yanase}, so that this tool is expected to work for
the examination whether superconductivity in iron pnictides, which are 
{\em moderately correlated}, can indeed be understood in 
terms only of FS nesting. 

Let us move on to the construction of the effective model
which we employ in the following calculation.
As for FLEX calculation for iron pnictides,
while there have already been several studies for
simplified models\cite{FLEX}, one of the present authors
recently performed a calculation for a five-orbital 
model based on LDA\cite{IkedaJPSJ}. 
There, he set the intra-orbital Coulomb interaction $U$
and the Hund coupling $J$ to be 1.44 eV and 0.24 eV, 
which are relatively small compared to those estimated by an
{\em ab initio} method\cite{cRPA1,cRPA2}.
In fact, 
it has been recognized that FLEX becomes problematic
for the five-orbital model
in the moderately correlated (realistic) regime
since some {\em uncontrollable} treatment is needed to get a 
reasonable spectral function.
Namely, we have to introduce an artificial 
level shift for $d_{z^2}$ to the original one-body 
Hamiltonian\cite{IkedaJPSJ}, otherwise 
the $d_{z^2}$ orbital makes a large FS 
around the $\Gamma$ point which is not consistent with
the experiments\cite{ARPES}. 

The reason why we have to shift the level of
$d_{z^2}$ can be understood as follows.
Since the the $d_{z^2}$ orbital in the five-orbital 
model is localized and correlated\cite{cRPA1}
(due to weak hybridization with As 4$p$),  
the correlation already taken 
in LDA is expected to be large for $d_{z^2}$. 
Thus, while we have to introduce the so-called double 
counting (DC) term\cite{DC} to combine LDA with FLEX, 
the DC term of $d_{z^2}$ should be larger than those of 
$d_{yz}$, $d_{xz}$ and $d_{x^2{\mathchar`-}y^2}$
(which make the FS in LDA).
However, it is a non-trivial task to evaluate the orbital 
dependent DC term from first principles for the five-orbital model,
especially when the system is moderately correlated.

\begin{figure}[ht]
\begin{center}
\includegraphics[width=7.5cm]{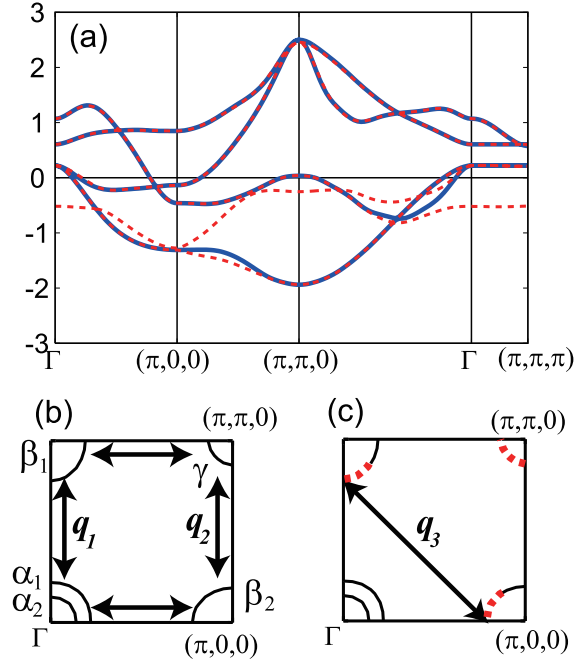}
\end{center}
\caption{(a)Band dispersion of the four-orbital model
(solid line) and the five-orbital model (dotted line) for LaFeAsO. 
(b)Fermi surface of the five-orbital model labeled as 
$\alpha_1$, $\alpha_2$, $\beta_1$, $\beta_2$ and $\gamma$, together
with nesting vectors ${\bf q_1}$ and ${\bf q_2}$.
(c)A plot similar to (b), where the 
$d_{x^2{\mathchar`-}y^2}$-character is dominant is shown by dotted lines.
}
\label{Fig1}
\end{figure}

In order to avoid this subtle problem and 
make our analysis on the validity of the nesting scenario
be clearer, in the present study, 
we first construct an effective four-orbital model for 
$d_{xy}$, $d_{x^2{\mathchar`-}y^2}$, $d_{yz}$, and $d_{xz}$.
As we can see in Fig.~\ref{Fig1}, since $d_{z^2}$ basically lies below 
the Fermi level, it is expected to be irrelevant for the 
physics of the FS nesting.
In actual model construction, as was done in Ref.~\cite{KurokiPRL},
we obtain the band structure 
of LaFeAsO with the Quantum-ESPRESSO package\cite{PWscf}, and
made maximally localized Wannier functions\cite{MaxLoc} 
for the four orbitals to represent the Kohn-Sham Hamiltonian 
in terms of these orbitals. 

In table \ref{tab1}, we list the transfer integrals in the
one-body part of the Hamiltonian. 
The onsite energies of $d_{yz}(d_{xz})$, $d_{x^2{\mathchar`-}y^2}$, 
$d_{xy}$ are 0.09, 0.28, and 0.11eV 
with respect to the Fermi level.
The spread of Wannier functions 
($\sqrt{\langle r^2 \rangle-\langle r \rangle^2}$)
are 2.16, 2.55 and 2.11 \AA, while in the five-orbital 
model\cite{Vildosola}, they are 2.05, 2.36, and 1.73 \AA,
respectively.

\begin{table*}[ht]
\begin{tabular}{c|rrrrrccc}
\backslashbox{\small($\mu$,$\nu$)}{\small $[\Delta X, \Delta Y]$}
            &[1,0]&[1,1]&[2,0]&[2,1]&[2,2]&$\sigma_Y$& $I$& $\sigma_d$\\ 
\hline
  ($zx$,$zx$)              &  $-$0.22&   0.18&       &        &        &+($yz$,$yz$)  & +   &+  \\
  ($zx$,$yz$)              &     0.12&       &   0.03&$-$0.02 &        &    +   & +   &$-$\\
  ($zx$,$x^2$-$y^2$)       &     0.20&       &       &   0.03 &        &+($yz$,$x^2$-$y^2$)  & $-$ &$-$\\
  ($zx$,$xy$)              &  $-$0.29&   0.19&       &  -0.02 &        &$-$($yz$,$xy$)& $-$ &+  \\
  ($yz$,$yz$)              &  $-$0.22&   0.30&       &        &   0.06 &+($zx$,$zx$)  & +   &+  \\
  ($yz$,$x^2$-$y^2$)       &     0.20&   0.03&       &        &        &+($zx$,$x^2$-$y^2$)  & $-$ &+  \\
  ($yz$,$xy$)              &     0.29&       &       &        &        &$-$($zx$,$xy$)& $-$ &$-$\\
  ($x^2$-$y^2$,$x^2$-$y^2$)&     0.15&   0.12&$-$0.06&$-$0.03 &        &   +    & +   &+  \\
  ($x^2$-$y^2$,$xy$)       &         &       &       &$-$0.05 &        &  $-$   & +   &$-$\\
  ($xy$,$xy$)              &     0.30&$-$0.20&   0.02&        &$-$0.03 &   +    & +   &+  \\
\end{tabular}
\caption{Hopping integrals $t(\Delta X, \Delta Y; \mu, \nu)$ 
in units of eV. $[\Delta X, \Delta Y]$ denotes the in-plain 
hopping vector (note that the $X$- and $Y$-axis point towards neighboring 
Fe atoms), and $(\mu,\nu)$ the orbitals. $\sigma_Y$, $I$, and 
$\sigma_d$ corresponds to $t(\Delta X, -\Delta Y;\mu,\nu)$, 
$t(-\Delta X, -\Delta Y;\mu,\nu)$, $t(\Delta Y, \Delta X;\mu,\nu)$, 
respectively, where `$\pm$' and `$\pm (\mu',\nu')$' in the row of 
$(\mu,\nu)$ mean that the corresponding  
hopping is equal to $\pm t(\Delta X, \Delta Y; \mu, \nu)$ and  
$\pm t(\Delta X, \Delta Y; \mu', \nu')$, respectively. 
This table, combined with the relation 
$t(\Delta X, \Delta Y; \mu, \nu) = t(-\Delta X, -\Delta Y; \nu, \mu)$, 
gives all the in-plain hoppings $\geq 0.02$eV up to the fifth 
neighbors.\label{tab1}}
\end{table*}

In Fig.\ref{Fig1}, we plot the band dispersion (in the
unfolded Brillouin zone\cite{KurokiPRL} for the unit cell which contains
only one Fe atom) of the four-orbital model, 
comparing with that of the five-orbital model\cite{comment}.
We can see that the FS is almost the same as that of the 
original five-orbital model. Thus we may expect that
if the nesting scenario indeed captures the essential feature of the
pairing mechanism, the four-orbital model should also be 
able to describe superconductivity in iron pnictides.

For the many-body
part of the Hamiltonian, we consider the standard
interactions that comprise the inter- and 
intra-orbital Coulomb interaction ($U$, $U'$)
and the Hund coupling and the pair-hopping ($J$, $J'$). 
We assume that $U=U'+2J$ and $J=J'$.
Following the standard procedures of multi-orbital 
FLEX\cite{KurokiPRB}, we solve the linearized
Eliashberg equation, and obtain the gap function 
along with the eigenvalue $\lambda$. $T_c$
corresponds to the temperature where $\lambda$ reaches unity.
32$\times$32$\times$4 three-dimensional $k$-point meshes 
and 1024 Matsubara frequencies are taken, while the temperature ($T$)
is set to be 0.02 eV. 
Hereafter we denote the spin susceptibility, the gap function,
and the squared absolute value of the Green function 
for the lowest Matsubara frequency ($i\omega_n$ = 0 for the
spin susceptibility and $i\pi T$ for the latter two)
and $k_z=0$ as $\chi_s(k_x,k_y)$, $\Delta(k_x,k_y)$, and 
$g(k_x,k_y)$, respectively.

\begin{figure}[ht]
\begin{center}
\includegraphics[width=8.5cm]{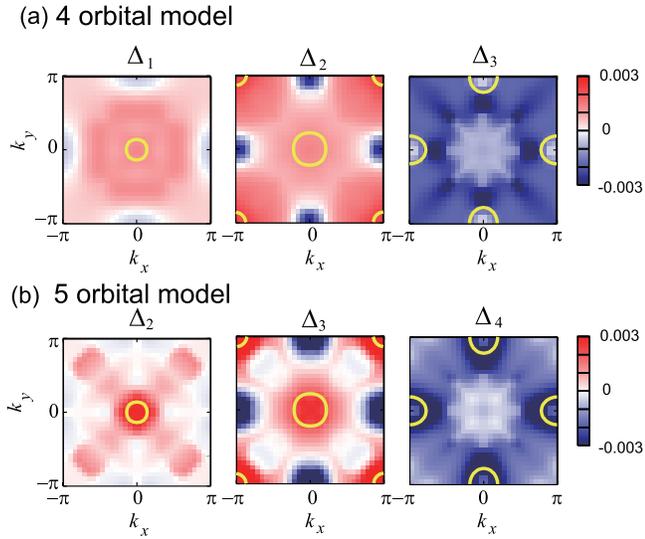}
\end{center}
\caption{
Gap functions for the four-orbital model (a)
(the five-orbital model (b)) of LaFeAsO,
for the 1st (2nd) to 3rd (4th) bands from bottom 
to top in the band representation with $n$ = 4.0 (6.0). 
Solid lines represent the Fermi surface.
}
\label{Fig2}
\end{figure}

First, we compare the four-orbital model and the five-orbital model
by RPA. Both models have a peak in the spin susceptibility 
which corresponds to stripe AF instability, 
and the fully-gapped $s$-wave with sign reversing
(the so-called $s_{\pm}$-wave) is most dominant
for the undoped case (electron-filling $n$ is 4.0 for the four-orbital
model and 6.0 for the five-orbital model). 
Interaction parameters are taken as $U=1.2$, $U'=0.9$, and $J=0.15$ eV. 
In Fig.~\ref{Fig2}, we plot $\Delta$ of the the four(five)-orbital 
model, for the 1st (2nd) to 3rd (4th) bands from 
bottom to top in the band representation. We see
that the gap functions of these two models are similar
to each other in RPA.

\begin{figure}[ht]
\begin{center}
\includegraphics[width=8.5cm]{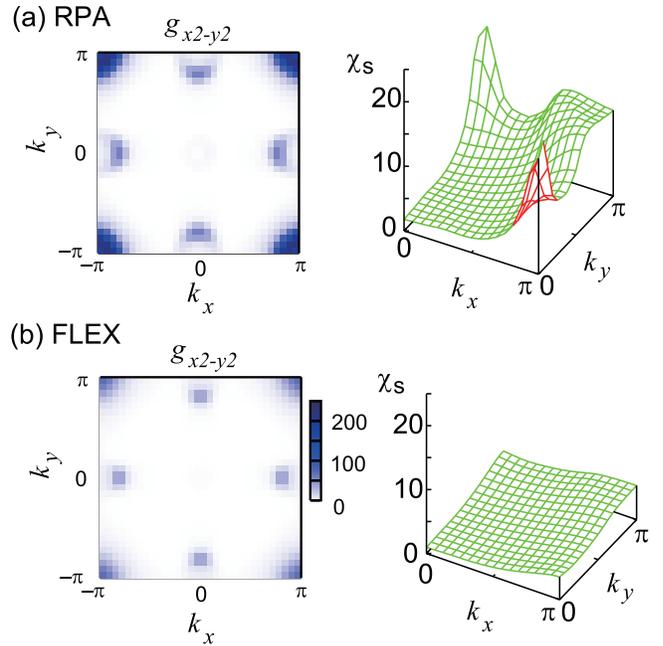}
\end{center}
\caption{
RPA(a) and FLEX(b) results for the squared
absolute value of the Green function for 
the $d_{x^2{\mathchar`-}y^2}$ orbital in the orbital representation
(left) and spin susceptibility at the lowest Matsubara
frequency (right) for the four-orbital model with $U=1.2$, $U'=0.9$,
and $J=0.15$ eV.}
\label{Fig3}
\end{figure}

Next let us move on to FLEX. In Fig.~\ref{Fig3},
we compare RPA and FLEX results of $g$ 
(which takes large values at FS, see also Fig.~\ref{Fig1}(c)) 
for $d_{x^2{\mathchar`-}y^2}$ in the orbital representation 
and $\chi_s$. Here we set $U=1.2$, $U'=0.9$ and $J=0.15$ eV.
Although we have a sharp peak in $\chi_s$ at $(k_x,k_y)=(\pi,0)$ 
and $(0,\pi)$ in RPA, these peaks are severely suppressed in FLEX
and a broad structure around $(\pi,\pi)$ becomes dominant.
While the fully-gapped $s_{\pm}$-wave
is dominant in both RPA and FLEX, $\lambda$ 
is much smaller in FLEX(=0.18) than in RPA(=0.83).
These results suggest that, in order to have strong
magnetic fluctuations and pairing instability from the
nesting between small Fermi pockets, generally we need
sufficiently large interactions.

\begin{figure}[ht]
\begin{center}
\includegraphics[width=8.5cm]{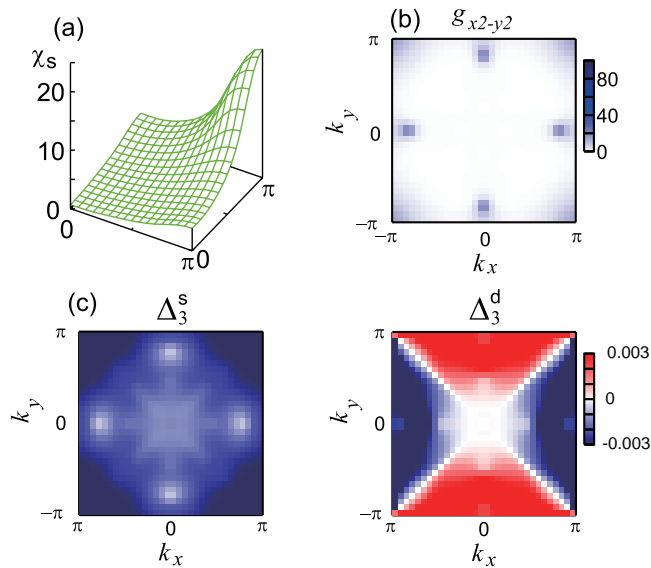}
\end{center}
\caption{
FLEX results for the four-orbital model with
$U=1.8$, $U'=1.2$, and $J=0.3$ eV. 
Spin susceptibility (a), squared absolute value of the
Green function for the $d_{x^2{\mathchar`-}y^2}$ orbital
(b) and the gap function of the $d_{x^2{\mathchar`-}y^2}$ orbital
for the $s_{\pm}$- and $d$-wave pairing
at the lowest Matsubara frequency (c) are shown.
}
\label{Fig5}
\end{figure}

In Fig.~\ref{Fig5}, we show the result of FLEX
for $U=1.8$, $U'=1.2$ and $J=0.3$.
In contrast with Fig.~\ref{Fig3}(b),
we have a sharp peak in $\chi_s$ (Fig.~\ref{Fig5}(a)).
However, 
the position of the peak is located at $(\pi,\pi)$ which
corresponds to the G-type AF instability.
Here, rather than the fully-gapped $s_{\pm}$-wave pairing, 
the so-called gapless $s_{\pm}$-wave pairing (whose
$|\Delta|$ becomes small on the Fermi surface\cite{KurokiPRB}) 
and a $d$-wave pairing are more favored ($\lambda$ is $\sim$ 0.34 
for these pairings, and see Fig.~\ref{Fig5}(c)
in which $\Delta$ of these pairings
for the third band in the band representation are plotted).
For larger interaction parameters, the G-type AF instability 
is more and more enhanced and the gapful $s_{\pm}$-wave 
pairing never becomes dominant.

These results can be understood as follows.
For undoped LaFeAsO, as we see in Fig.~\ref{Fig1}(b) and (c),
there are four kinds of Fermi pockets:
two hole pockets, $\alpha_1$ and $\alpha_2$, around $(0,0)$  
and electron pockets, $\beta_1$, $\beta_2$, around
$(\pi,0)$ and $(0,\pi)$
as well as a hole pocket $\gamma$ around $(\pi, \pi)$ 
(which disappears by $\sim 10\%$ doping of electrons).
While $\alpha$ is made from 
$d_{yz}$and $d_{xz}$, $\beta$ is made from 
$d_{yz}$, $d_{xz}$ and $d_{x^2{\mathchar`-}y^2}$, and $\gamma$ is
made from $d_{x^2{\mathchar`-}y^2}$ (Fig.~\ref{Fig1}(c)).
On top of the scattering channels which connect $\alpha$ and $\beta$
or $\gamma$ and $\beta$ (${\bf q_1}$ and 
${\bf q_2}$ in Fig.~\ref{Fig1}(b)), there is also
a channel between $d_{x^2{\mathchar`-}y^2}$ in $\beta_1$
and $\beta_2$ (${\bf q_3}$ in Fig.~\ref{Fig1}(c)). 
As is discussed in 
Ref.~\cite{KurokiPRB}, while ${\bf q_2}$ is the key to
induce the strong stripe AF fluctuations,
in the absence of $\gamma$, 
${\bf q_3}$ suppresses the stripe AF fluctuations
and the fully-gapped $s_{\pm}$-wave pairing.  
In fact, the low energy electronic structure in FLEX is quite similar to
that in RPA: the spectral functions obtained by 
these approximations have similar size of
structures in the Brillouin zone around
$(\pi,0)$, $(0,\pi)$, $(0,0)$ and $(\pi,\pi)$ for low frequencies. 
However, the self-energy effect for $d_{x^2{\mathchar`-}y^2}$ 
around $(\pi,\pi)$ (where the LDA band dispersion is flat)
is significantly large 
and severely suppresses ${\bf q_2}$, which is dominant in RPA.
While effects which have not been considered 
(such as spin-lattice coupling, off-site Coulomb interactions, etc.)
may change the situation, the present study
suggests that the nesting scenario is not necessarily robust
against the momentum-dependent self-energy correction.

To summarize, we study whether 
FS nesting can be the origin of spin-fluctuation-mediated
superconductivity in iron-pnictides by means of FLEX.
Our result suggests that the nesting scenario has a serious dilemma: 
since all the FS pockets are small, we need sufficiently strong 
interactions to have large magnetic fluctuations from the FS nesting. 
However, due to the self-energy corrections, the contribution of
the low-energy $d_{x^2{\mathchar`-}y^2}$ 
state at $(\pi,\pi)$ which is the key to make dominant stripe
AF fluctuations in the weak coupling regime is
severely suppressed for the moderately 
correlated regime.  The remaining $d_{x^2{\mathchar`-}y^2}$ 
states at $(0,\pi)$ and $(\pi,0)$ induce the G-type AF fluctuations
which favor the $d$- or nodal $s$-wave pairing
and the high $T_c$ nodeless $s_{\pm}$-pairing is difficult to arise.

\section*{Acknowledgment}
We would like to thank H. Aoki, P. Hansmann, M. Imada, 
G. Sangiovanni, A. Toschi for fruitful discussions
and/or critical reading of the manuscript.
This work is supported by a Grant-in-Aid for Scientific Research
on Priority Areas ``Super Clean Materials'' 
under Grant No. 20029014 from MEXT, Japan. 
Numerical calculations were performed at
the facilities of the Supercomputer center,
ISSP, University of Tokyo.


\begin{thebibliography}{99}
\bibitem{comment} It should be noted that the band dispersion of the
five-orbital model is exactly the same as that of ab-initio calculation
\cite{KurokiPRL,cRPA1}.
\end{thebibliography}
\end{document}